
\newskip\oneline \oneline=1em plus.3em minus.3em
\newskip\halfline \halfline=.5em plus .15em minus.15em
\newbox\sect
\newcount\eq
\newbox\lett
\newdimen\short
\def\adv{\global\advance\eq by1}
\def\set#1#2{\setbox#1=\hbox{#2}}
\def\nextlet#1{\global\advance\eq by-1\setbox
                \lett=\hbox{\rlap#1\phantom{a}}}

\newcount\eqncount
\newcount\sectcount
\eqncount=0
\sectcount=0
\def\sectadv{\global\advance\sectcount by1    }
\def\secta{\global\advance\sectcount by1    }
\def\equn{\global\advance\eqncount by1\eqno{(\copy\sect.\the\eqncount)} }
\def\put#1{\global\edef#1{(\the\sectcount.\the\eqncount)}}
\def\puu#1{\global\edef#1{(\copy\sect.\the\eqncount)}}

\magnification = 1200{\rm}

\voffset 0.2 truein
\vsize   8.7 truein
\hsize  6.25 truein
\hfuzz 30 pt

\def\mbox#1#2{\vcenter{\hrule \hbox{\vrule height#2in
                \kern#1in \vrule} \hrule}}  
\def\sq{\,\raise.5pt\hbox{$\mbox{.09}{.09}$}\,}
\def\sqb{\,\raise.5pt\hbox{$\overline{\mbox{.09}{.09}}$}\,}
\def\tri{\triangle}

\def\R{R^2}
\def\Ric{R_{ab}R^{ab}}
\def\Rie{R_{abcd}R^{abcd}}

 2

\def\a{\alpha}
\def\b{\beta}

\def\d{\delta}
\def\e{\epsilon}
\def\f{\phi}

\def\g{\gamma}
\def\h{\eta}

\def\k{\kappa}
\def\l{\lambda}
\def\m{\mu}
\def\n{\nu}
\def\o{\omega}
\def\p{\pi}

\def\r{\rho}
\def\s{\sigma}
\def\t{\tau}

\def\x{\xi}
\def\z{\zeta}

\def\G{\Gamma}

\def\O{\Omega}

\def\cc{{\cal C}}
\def\cd{{\cal D}}

\def\ck{{\cal K}}

\def\cm{{\cal M}}

\def\co{{\cal O}}

\def\cy{{\cal Y}}

\def\pa{\partial}
\def\de{\nabla}

\def\TH{{\raise.2ex\hbox{$\displaystyle \bigodot$}\mskip-4.7mu \llap H \;}}
\def\face{{\raise.2ex\hbox{$\displaystyle \bigodot$}\mskip-2.2mu \llap {$\ddot
        \smile$}}}

\def\Hat#1{\rlap{\kern.10em$\widehat{\phantom G}$}#1}
\def\HAt#1{\rlap{\kern.05em$\widehat{\phantom G}$}#1}

\def\cap#1{\rlap{\kern.1em$\widehat{\phantom{G\vrule height.8em}}$}#1{}}
\def\Cap#1{\rlap{\kern.05em$\widehat{\phantom{G\vrule height.8em}}$}#1{}}

\def\leftrightarrowfill{$\mathsurround=0pt \mathord\leftarrow \mkern-6mu
        \cleaders\hbox{$\mkern-2mu \mathord- \mkern-2mu$}\hfill
        \mkern-6mu \mathord\rightarrow$}
\def\overleftrightarrow#1{\vbox{\ialign{##\crcr
        \leftrightarrowfill\crcr\noalign{\kern-1pt\nointerlineskip}
        $\hfil\displaystyle{#1}\hfil$\crcr}}}

\def\frac#1#2{{\textstyle{#1\over\vphantom2\smash{\raise.20ex
        \hbox{$\scriptstyle{#2}$}}}}}
\def\ha{\frac12}

\catcode`@=11
\def\underline#1{\relax\ifmmode\@@underline#1\else
        $\@@underline{\hbox{#1}}$\relax\fi}
\catcode`@=12

\def\nis{\nointerlineskip}
\def\Abar{\vbox{\nis\moveright.33em\vbox{
        \hrule width.35em height.04em}\nis\kern.05em\hbox{$A$}}{}}
\def\Dbar{\vbox{\nis\moveright.20em\vbox{
        \hrule width.50em height.04em}\nis\kern.05em\hbox{$D$}}{}}
\def\Gbar{\vbox{\nis\moveright.20em\vbox{
        \hrule width.50em height.04em}\nis\kern.05em\hbox{$G$}}{}}
\def\mbar{\vbox{\nis\moveright.15em\vbox{
        \hrule width.60em height.04em}\nis\kern.05em\hbox{$m$}}{}}
\def\Rbar{\vbox{\nis\moveright.20em\vbox{
        \hrule width.50em height.04em}\nis\kern.05em\hbox{$R$}}{}}
\def\Vbar{\vbox{\nis\moveright.05em\vbox{
        \hrule width.60em height.04em}\nis\kern.05em\hbox{$V$}}{}}
\def\Xbar{\vbox{\nis\moveright.20em\vbox{
        \hrule width.60em height.04em}\nis\kern.05em\hbox{$X$}}{}}
\def\thetabar{\vbox{\nis\moveright.15em\vbox{
        \hrule width.30em height.04em}\nis\kern.05em\hbox{$\theta$}}{}}
\def\Lambdabar{\vbox{\nis\moveright.25em\vbox{
        \hrule width.35em height.04em}\nis\kern.05em\hbox{${\mit\Lambda}$}}{}}
\def\Sigmabar{\vbox{\nis\moveright.25em\vbox{
        \hrule width.50em height.04em}\nis\kern.05em\hbox{${\mit\Sigma}$}}{}}
\def\phibar{\vbox{\nis\moveright.18em\vbox{
        \hrule width.40em height.04em}\nis\kern.05em\hbox{$\phi$}}{}}
\def\chibar{\vbox{\nis\moveright.12em\vbox{
        \hrule width.40em height.04em}\nis\kern.05em\hbox{$\chi$}}{}}
\def\psibar{\vbox{\nis\moveright.23em\vbox{
        \hrule width.40em height.04em}\nis\kern.05em\hbox{$\psi$}}{}}
\def\debar{\vbox{\nis\moveright.18em\vbox{
        \hrule width.35em height.04em}\nis\kern.05em\hbox{$\partial$}}{}}
\def\delbar{\vbox{\nis\moveright.10em\vbox{
        \hrule width.63em height.04em}\nis\kern.05em\hbox{$\nabla$}}{}}

\def\Y{Y_{_{JM}}}
\def\Ys{Y^*_{_{JM}}}
\def\Yh{Y_{\ha M}}
\def\dT{{\dot{T}}_{JM}^{(+)}}
\def\ph{\varphi_{_{JM}}}
\def\phd{\varphi^{\dag}_{_{JM}}}
\def\jo{J_1}
\def\jt{J_2}
\def\mo{M_1}
\def\mt{M_2}


\def\cFIG{1}
\def\cAntMot{2}
\def\cAMM{3}
\def\cGSW{4}
\def\cCT{5}
\def\cBD{6}
\def\cWDW{7}
\def\cVar{8}

\nopagenumbers
\rightline{CPTH-S376.0995}
\rightline{LA-UR-95-3392}
\rightline{hep-th/9509168}
\rightline{September 24, 1995}
\vskip 1.2truecm
\centerline{\bf {QUANTUM DIFFEOMORPHISMS AND CONFORMAL SYMMETRY}}
\vskip 1.2truecm
\centerline{Ignatios Antoniadis}
\centerline{{\it Centre de Physique Theorique}
\footnote{$^{\dag}$}{\it Laboratoire Propre du CNRS UPR A.0014}}
\centerline{{\it Ecole Polytechnique}}
\centerline{{\it 91128 Palaiseau, France}}
\vskip .5truecm
\centerline{Pawel O. Mazur}
\centerline{{\it Dept. of Physics and Astronomy}}
\centerline{{\it University of South Carolina}}
\centerline{{\it Columbia, SC 29208 USA}}
\vskip .3truecm
\centerline{and}
\vskip .3truecm
\centerline{Emil Mottola}
\centerline{{\it Theoretical Division, T-8}}
\centerline{{\it Mail Stop B285}}
\centerline{{\it Los Alamos National Laboratory}}
\centerline{{\it Los Alamos, NM 87545 USA}}
\vskip 1truecm
\centerline{\bf Abstract}
\vskip .5truecm
We analyze the constraints of general coordinate invariance
for quantum theories possessing conformal symmetry in four dimensions.
The character of these constraints simplifies enormously on the
Einstein universe $R \times S^3$. The $SO(4,2)$ global conformal
symmetry algebra of this space determines uniquely a finite shift
in the Hamiltonian constraint from its classical value. In other words,
the global Wheeler-De Witt equation is {\it modified} at the quantum
level in a well-defined way in this case. We argue that the higher
moments of $T^{00}$ should not be imposed on the physical states
{\it a priori} either, but only the weaker condition
$\langle \dot T^{00} \rangle = 0$. We present an explicit example
of the quantization and diffeomorphism constraints
on $R \times S^3$ for a free conformal scalar field.

\vfill
\eject

\baselineskip=20pt

\footline={\hss\tenrm\folio\hss}
\pageno = 1
\newcount\eqncount
\sectadv
\set\sect{1}
\beginsection{1. Introduction}

The equality of gravitational and inertial mass led Einstein to the unshakable
conviction that any theory of gravitation must respect the
Equivalence Principle. General coordinate invariance is the
most elegant mathematical expression of the Equivalence Principle, and
consequently
it was to play a central role in the development of classical general
relativity.
In view of the importance of general coordinate invariance in the classical
theory,
and in the absence of any evidence that such a pivotal guiding principle
should be
abandoned {\it a priori}, it is natural to regard diffeomorphism invariance
as an
essential component of any proposal for a quantum theory of gravity.

Unfortunately, although the Einstein theory is
certainly coordinate invariant classically, the status of this
important invariance becomes obscured in the traditional approaches to
quantization. If we look beyond the myriad of technical difficulties in
these approaches such as ordering ambiguities, choice of gauge, and the
ill-defined
nature of operators at the same spacetime point, the basic problem at their
root is
that there is no unique extension of the algebra of diffeomorphisms
off-shell. As a
result, the invariance or non-invariance of the theory becomes intertwined
with the
time evolution dynamics in a completely non-trivial way. Since it is virtually
certain that the Einstein theory is only an effective theory which must be
modified
in order to construct a consistent quantum generalization, it is a great
handicap to
a wider exploration of possible quantum extensions to embed coordinate
invariance
so inextricably into the context of classical general relativity. Without a
clear
expression of the invariance principle in a mathematically consistent framework
independent of dynamical assumptions, one cannot even ``get off the ground" in
quantum gravity.

It is largely for this reason of formulating diffeomorphism invariance
independently of detailed dynamics that the functional integral approach
to quantization offers distinct advantages [\cFIG]. There one can see a clean
separation of the problem into first determining the invariant integration
measure on the space of metrics and then the action functional to be
integrated. It
is from the invariant path integral measure that one may extract most
directly the
important consequences of general coordinate invariance in the quantum
theory. In
particular, the introduction of the conformal parameterization of the metric,
$$
g_{ab}(x) = e^{2 \s(x)} \bar g_{ab}(x), \equn\put\confdef
$$
leads quickly to the general form of the trace anomaly for
matter fields in a curved background, and the effective action
it induces for the $\s$ part of the metric [\cAntMot].
The Faddeev-Popov ghost determinant arises from the change
of integration variables represented by {\confdef}, and it
takes a form analogous to that in two dimensional
quantum gravity ({\it i.e.} non-critical string theory) [{\cFIG}-{\cAMM}].
The conformal sector of quantum gravity in four dimensions
may be studied then by methods closely paralleling those in two
dimensions, where a great deal of progress has been made in the
last decade. One consequence of this approach (in both two and four
dimensions) is that although conformal symmetry is broken by the
trace anomaly in a fixed background, conformal invariance is recovered
when the quantum $\s$ theory is considered. In fact, the quantum
$\s$ theory describes a conformal fixed point of quantum gravity
where scale invariance is restored and the total trace anomaly
of matter plus $\s$ plus ghosts vanishes [\cAMM].

These results strongly suggest that
scale invariance should play an important role in the full quantum
theory. Global scale transformations are in any case just particular
coordinate transformations of any spacetime that is conformally
flat, of which the Friedman-Robertson-Walker metric of classical cosmology is a
prime example. Hence conformal invariance should play a important role in the
description of quantum gravitational effects in our  Universe as well.

Motivated by these considerations we study in this paper the
constraints of coordinate invariance in theories possessing
global conformal symmetry. Though powerful and elegant, the functional
integral approach leaves unanswered the question of the algebra of
diffeomorphisms off-shell, the Hilbert space of states, and
the physical states of any theory invariant under diffeomorphisms.
Addressing these issues requires a canonical operator approach to
quantization. By requiring that the trace of the total
energy-momentum tensor vanishes and quantizing in the conformally flat
Einstein universe $R \times S^3$, we shall show that the constraints
of diffeomorphism invariance vastly simplify and a detailed
analysis of the constraints in four dimensions suddenly becomes possible.
Otherwise
we do not need to specify the Lagrangian nor supply any other dynamical
information
of the theory. Hence our framework and the results are quite independent of the
Einstein theory (although it too is conformally invariant  {\it on-shell} in
the
trivial sense that $T_a^a = 0$ follows by tracing the equations of motion).

The technical advantage of quantizing on the conformally flat spacetime
$R\times S^3$ is that it admits a complete classification of states,
labeled by the integers, according to their weight under conformal
transformations. Hence, the constraint conditions may be solved level by level
in a straightforward algebraic manner. On $R \times S^3$ rigid
conformal transformations are contained explicitly in the set of volume
non-preserving reparametrizations of the spatial $S^3$. Hence global
conformal invariance is imposed as a direct consequence of diffeomorphism
invariance, and conformal transformations are treated together with and on the
same footing as any other coordinate reparametrization under which the
physical spectrum must be invariant. Moreover, the global conformal algebra
$SO(4,2)$ on $R \times S^3$ contains the quantum Hamiltonian as one of its
generators, and the conformal weights correspond to the eigenvalues of the
Hamiltonian. Because $R\times S^3$ is a product space, the time translations
generated by the Hamiltonian are not mixed with spatial reparametrizations and
there are no operator ordering ambiguities. This is an important feature of the
approach since it allows us to determine the quantum Hamiltonian constraint
uniquely and in a completely unambiguous manner.

In $D=2$ when quantizing the Liouville theory on $R\times S^1$, the $L_0$
condition on the physical states is modified by a finite c-number shift,
which arises from the $bc$ ghost action needed for quantization in covariant
gauges such as {\confdef} [\cGSW]. We show by explicit construction that a
similar
well-defined shift in the Hamiltonian condition on the physical states
occurs in
$D=4$. The source of this shift lies again in the gauge fixing procedure which
results in a modification of the classical Lie algebra of the finite
dimensional
global conformal group $SO(4,2)$ at the quantum level in the ghost sector, {\it
i.e} it arises from the covariant integration measure  of the path integral for
quantum gravity.

This modification of the classical
Hamiltonian constraint by the quantum measure is important because it
demonstrates explicitly that the classical Hamiltonian constraint
{\it cannot} be maintained at the quantum level in general. Hence all
attempts to implement the classical constraints of general coordinate
invariance without attention to the algebra of diffeomorphisms at
the quantum level must be viewed as questionable. In particular,
in the canonical approach on $R \times S^3$ one does not enforce
the strong form of the classical time reparametrization constraint,
$T^{00} = 0$. Instead only the considerably weaker condition
$\langle \dot T^{00} \rangle = 0$ should be imposed on matrix elements
between physical states, in analogy with the procedure in $D=2$.

That strong vanishing of $T^{00}$ cannot be imposed as an operator
condition is clear enough, since it must have non-zero commutators
in order to generate time reparametrizations in the quantum theory.
That only the positive frequency component of $T^{00}$ should be
required to vanish on the physical (ket) states is a consequence of the
Dirac-Fock approach to the quantization of theories with constraints
that we follow. Then the vanishing of the matrix elements of $\dot T^{00}$
follows immediately from the fact that the negative frequency part $T^{00(-)}$
(being the Hermitian conjugate of $T^{00(+)}$) annihilates the physical
(bra) states
acting to the left. However, there is {\it no} such condition on the zero
frequency
part which remains unfixed by this procedure, except for its spatial
integral which
is the Hamiltonian. Any condition on the full
$T^{00}$ should emerge from the {\it dynamics} of the theory under
time evolution, and not be imposed as a kinematical constraint in the
canonical approach.

The outline of the paper is as follows. In Section 2 we define the problem
of the diffeomorphism constraints and classical Lie algebra of the global
conformal group on $R\times S^3$, introducing necessary notation and
properties of the spherical harmonic functions on $S^3$ which we
require in the subsequent analysis. In Section 3 we discuss the
modification of the
classical algebra arising from the quantum measure, and construct the finite
dimensional $bc$ ghost action whose quantization leads to the c-number
shift in the
quantum Hamiltonian constraint. In section 4 we derive detailed expressions
for the
moments of the $T^{0i}$ constraints on $S^3$ for a massless, conformally
coupled
scalar field, illustrating the algebra of spatial diffeomorphisms at the
quantum
level in a specific example.   We show that the surviving physical state of the
scalar has a correspondence with an operator of well-defined scaling
dimension in
four dimensions. Some properties of the harmonics on $S^3$ needed in the
analysis
are given in the Appendix for completeness.

This paper is paper I of two. In paper II we apply the general method
of this paper to the analysis of the diffeomorphism constraints of
the effective theory of the conformal factor generated by the
trace anomaly.

\newcount\eqncount
\sectadv
\set\sect{2}
\beginsection{2. Diffeomorphism Constraints on $R \times S^3$}

We follow the parallel of
radial quantization in two dimensional conformal field theories and choose
as our background spacetime the static Einstein universe $R \times S^3$
with metric,
$$
\bar g_{ab}\, dx^a\,  dx^b = -dt^2 + a^2\, d\O^2\ ,\equn\put\bkg
$$
where $d\O^2$ is the standard round metric on the unit three sphere.
In what follows we shall set the radius, of $S^3$ equal to unity, $a=1$
and drop the overbar on the metric background where it causes no confusion
to do so. This metric background is conformally related to ordinary Lorentzian
flat spacetime in such a way that time translations in the static Einstein
space correspond to global dilations of flat Minkowski space. This
is most easily demonstrated by first continuing to Euclidean signature
$t \rightarrow -i\t$:
$$
d\t^2 + d\O^2 = e^{2\t}(dr^2 + r^2 d\O^2) = e^{2\t}d s_{flat}^2 \ ,
\equn\put\cmap
$$
with
$$
r =e^{-\t}\ . \equn\put\trans
$$
This is very convenient for our purposes since it implies that the spectrum
and eigenstates of the Hamiltonian operator on $R \times S^3$ are in one-to-one
correspondence with the spectrum of scaling dimensions in the conformal  field
theory description. This is analogous to two dimensional gravity  where
canonical
quantization was carried out on the cylinder $R \times S^1$ [\cCT].

The diffeomorphisms are generated by the energy-momentum tensor $T^{ab}$.
Given any coordinate invariant action the purely spatial
diffeomorphism on the sphere $\xi_i$ is generated by the $T^{0i}$ components
of the energy-momentum tensor via
$$
X_{\xi} \equiv - \int_{S^3}\, d \O\, \xi_i\,T^{0i}\ .\equn
$$
The commutator of two such spatial diffeomorphisms on $S^3$ is also a
spatial diffeomorphism, {\it i.e.}
$$
[iX_{\x}, iX_{\z}] = iX_{[\x , \z]} \ , \equn\put\spat
$$
where $[\x , \z]$ is the classical Lie bracket,
$$
[\x ,\z]^a = \z^b\de_b\x^a - \x^b\de_b\z^a \ ,\equn\put\lbrdef
$$
in the coordinate representation. This relation must remain true at the quantum
level in order for the algebra of spatial diffeomorphisms to remain free of
anomalies, a property we shall check {\it a posteriori}. The fact that the
algebra
of purely spatial diffeomorphisms closes upon itself is essentially a
statement of
kinematics in that it must be true independently of the time evolution or
detailed
dynamics of the theory.

Spatial diffeomorphisms may be divided into two classes: those which
preserve the
volume and those which do not preserve it. The volume preserving
diffeomorphisms
are generated by divergence-free transverse vectors,
$\nabla_i \x^i = 0$ and form a subalgebra of all spatial diffeomorphisms. On
the
other hand the volume non-preserving ones are generated by vectors which
are pure
gradients, $\x_i =\de_i \phi$, and do not form a closed subalgebra since their
commutators yields through their Lie bracket also the volume preserving ones.

In contrast to the $T^{0i}$ which generate spatial diffeomorphisms, the
$T^{00}$
component generates time reparametrizations which is intertwined with dynamics,
and its commutator with $T^{0i}$ gives rise to the space-space components
$T^{ij}$.
Indeed in flat spacetime the components of the energy-momentum tensor obey the
classical current algebra [\cBD]
$$
[T^{00}(t, \vec x), T^{0i}(t, \vec x')]_{flat} = -i\left( T^{ij}(\vec x) +
T^{00}(\vec x') \d^{ij} \right)\pa_j \d^3 (\vec x -\vec x') \ .\equn\put\curr
$$
The brackets of the spatial components $T^{ij}$ do not form a closed algebra in
general. Rather the $T^{ij}$ contain information about model dependent dynamics
over and above the constraints of diffeomorphism invariance.

The lowest moment of $T^{00}$ is simply the Hamiltonian
$$
H \equiv -\int_{S^3}\, d\O \h_{a}\, T^{0a} = \int_{S^3}\, d \O \,T^{00}\
\equn\put\ham
$$
which generates rigid time translations in the product space {\bkg},
corresponding
to the constant Killing vector
$$
\h^a = -\h_a = (1,\vec 0) \equn\put\time
$$
We note from the classical current algebra {\curr} that in the special case of
this lowest moment of the $T^{00}$ (and only for this moment),
the Lie bracket again closes on the $T^{0i}$ since $H$ generates the time
evolution via
$$
[H, X_{\x}] = -i {\partial X_{\x}\over \partial t}\ .\equn
$$

Classically demanding that the theory be invariant under both
space and time coordinate transformations implies $T^{0i} = T^{00} = 0$.
These constraints are automatically satisfied for any classical metric theory
of gravity since the equations of motion are $T^{ab} = 0$.
When we pass from the classical to the quantum theory the energy-momentum
tensor becomes an operator and the Lie bracket becomes a commutator. We cannot
impose the constraint of vanishing $T^{0i}$ as an operator condition
because of its
non-trivial commutators {\spat} and {\curr}. As in Dirac's approach to the
Gauss
Law constraint in electrodynamics or the quantization of two dimensional
gravity,
our strategy will be to demand instead that the generators $X_{\x}$ of the
spatial
diffeomorphism symmetry at a fixed time vanish {\it weakly} on the physical
states.
In other words, we shall require that all matrix elements of the spatial
diffeomorphism generators between physical states vanish, although the
operators
themselves do not vanish. A sufficient condition for the weak vanishing of the
matrix elements of the diffeomorphism constraints is that the {\it positive
frequency} part $X^{(+)}_{\xi}$ of the generators annihilate the physical
states,
{\it i.e.}
$$
X^{(+)}_{\xi}\vert phys\rangle = 0\ .\equn\put\Xphys
$$
This amounts to imposing invariance of the quantum theory under
infinitesimal spatial reparametrizations. The unambiguous separation of
$X_{\xi}$
into positive and negative frequency components is possible only on a
product space
such as $R \times S^3$ possessing the timelike Killing vector {\time}. The zero
frequency components must be treated separately.

Because the commutator of $T^{00}$ does not form a closed algebra with the
$T^{0i}$
off-shell, but instead brings in other quantities which involve dynamics,
we {\it
cannot} impose the local Wheeler-De Witt condition that $T^{00}$ (and all of
its
moments) vanish on the physical states [\cWDW]. Moreover, we should expect
that the
classical Lie bracket in {\curr} is generally modified at the quantum
commutator
level by anomalies. The well-known trace anomaly is the simplest example of
these
modifications to the classical algebra which must be taken into account in the
quantum theory. Far from being kinematic in nature the exact form of such
modifications to the classical algebra depends upon dynamics, and is therefore
non-trivial. Hence the $T^{00}$ and $T^{ij}$ cannot be imposed on the physical
states {\it a priori} without in effect knowledge of the full quantum
theory right
from the outset.

However, most of the classical time reparametrization invariance is recovered
in
the quantum theory by the imposition of the volume non-preserving spatial
diffeomorphism constraints since
$$
\int_{S^3} d\O\, (\nabla_i \phi)\, T^{0i} = \partial_t \int_{S^3}d\O\, \phi\,
T^{00}\equiv \dot T_{\phi}\ , \equn\put\clal
$$
for time-independent $\phi$.
By Fourier transforming in time it is clear that this relation together with
{\Xphys} guarantees the weak vanishing of all the matrix elements of $T^{00}$
containing non-zero frequencies,
$$
\langle phys' \vert \dot T_{\phi} \vert phys \rangle = 0 \ .\equn\put\phys
$$
Hence, not the Wheeler-De Witt condition of vanishing $T^{00}$ but the weaker
condition of vanishing matrix elements of $\dot T^{00}$ is imposed on the
physical
Hilbert space of states. The zero frequency component of $T^{00}$ is {\it not}
fixed by this condition. Finally, the lowest moment of the zero frequency
component, the global Hamiltonian constraint of vanishing $H$ also should not
be
imposed uncritically on the physical states. We shall see precisely how it
should be
modified by a finite subtraction in the quantum treatment of the next section.

It is instructive to compare and contrast this situation in four dimensions
with two
dimensions, on $R \times S^1$ [\cGSW]. In $D=2$ the space-space component
$T^{11}$
is determined by $T^{00}$ and the trace of the energy-momentum tensor, which is
zero classically for conformally invariant matter. This means that there
are no new
operators in the spatial components and the full energy-momentum algebra does
close. In fact, the moments of
$T^{00}$ and $T^{01}$ are
$$\eqalign{
T^{00}_{_N} &= L_{_N}e^{-iNt} + \bar L_{_{-N}}e^{iNt}\cr
T^{01}_{_N} &= L_{_N}e^{-iNt} - \bar L_{_{-N}}e^{iNt}\ ,\cr}
\equn\put\Ttwo
$$
where $L_{_N}$ and $\bar L_{_{-N}}$ are the Virasoro generators for left
and right movers on the circle $S^1$.
Quantum mechanically there is an anomaly which appears only in the form of a
c-number central extension of the classical algebra {\curr}. In fact, the
Virasoro
algebra,
$$
\left[L_{_M}, L_{_N}\right] = (M-N) L_{_{M+N}} +
{c\over 12}(M^3 - M)\d_{_{M,-N}}
$$
for both the left and right movers implies (at $t=0$),
$$\eqalign{
\left[ T^{01}_{_M} ,T^{01}_{_N}\right] &= (M-N) T^{01}_{_{M + N}}\qquad
{\rm but}\cr
\left[ T^{00}_{_M} ,T^{01}_{_N}\right] &= (M-N) T^{00}_{_{M+N}} +
{c\over 6}(M^3 - M)\d_{_{M,-N}}\ .\cr}
\equn\put\twomod
$$
Therefore, the purely spatial quantum algebra is identical to the classical
Lie algebra {\spat} and free of any anomaly, but the second commutator
{\twomod}
acquires a quantum correction with the central charge $c$ proportional to
the trace
anomaly coefficient.  The resulting algebra of both space and time
reparametrizations is a local conformal algebra which has non-trivial
representations in conformal field theories. This local Virasoro algebra is
also
isomorphic to that obtained from the positive and negative frequency
spatial moments
($L_{_N}$ and $\bar L_{_N}$ with $N$ positive or negative respectively) of the
$T^{01}$ only. Moreover, the Hamiltonian operator $H$ ($=L_0 +\bar L_0$ in
$D=2$)
requires a finite subtraction. In $D=4$ there is no infinite dimensional local
conformal symmetry, but there is still a global conformal group of which
$H$ is one
generator, which results from the separation of the moments of $T^{0i}$ into
positive and negative frequencies, as we now show.

The symmetry group of the spatial $S^3$ is
$O(4)\cong SU(2) \times SU(2)/Z_2$, and we shall make extensive use of this
symmetry group in our development. The finite dimensional representations of
$SU(2) \times SU(2)$ are characterized by two angular momentum quantum numbers
$(J_1,J_2)$ with $J_1$ and $J_2$ taking on integer or half-integer values. The
scalar spherical harmonics on $S^3$ belong to the $(J, J)$ representation
of $SU(2)
\times SU(2)$. We denote these scalar harmonic functions on $S^3$ by $Y_{Jmm'}$
where $m$ and $m'$ are the magnetic quantum numbers of the two $SU(2)$
subgroups, or
more compactly by $Y_{JM}$ with $M\equiv (m,m')$ denoting the pair of $SU(2)$
magnetic indices taking $(2J+1)^2$ distinct values in total. Explicit
representations for the $\Y$ will not be needed, but are given for
completeness in
the Appendix in terms of standard $SU(2)$ transformation Wigner $D$ functions,
where further properties of these functions are catalogued as well [\cVar].

Because $R \times S^3$ is conformally flat, it has fifteen conformal
Killing vectors  $\o_a$, satisfying the conformal Killing equation,
$$
(L \o)_{ab} \equiv \de_{a} \o_{b} + \de_{b} \o_{a} - \ha
g_{ab} \de^{c} \o_{c} = 0 \ ,
\equn\put\ckv
$$
the maximal number for a four dimensional manifold, and the same number as
flat Minkowski spacetime. Corresponding to each conformal Killing vector
there is a Noether charge,
$$
\int_{S^3} d\O\, \o_a T^{0a}
$$
which is time-independent because of the conservation of the current,
$\o_bT^{ab}$ upon using {\ckv} and $T_a^a=0$.

The six purely spatial Killing vectors of $S^3$
(which are conformally mapped into the three rotations and three boosts of
Minkowski spacetime) clearly satisfy this equation. These six solutions
of {\ckv} have vanishing time component and may be denoted as $\r= (0,\r^i)$
where the spatial index $i$ runs from $1$ to $3$. It is not difficult to find
an
explicit representation of the $\r^i$ in terms of the $\Y$ harmonic functions
on $S^3$:
$$
\r^i_{_{\mo\mt}} = i\,\hbox{$V \over 4$}\, Y^*_{{1\over 2} M_1}
{\buildrel\leftrightarrow\over{\de}}\, ^i\, Y_{{1\over 2} M_2}
\equn\put\ks
$$
where $V = 2\p^2$ is the volume of the unit $S^3$. This
anti-symmetrized form makes it clear that
$$
\de_i\,\r^i_{_{\mo\mt}} = 0\quad {\rm and}
\quad\left[\rho^i_{_{\mo\mt}}\right]^*=
\rho^i_{_{M_2M_1}}
\equn\put\rrel
$$
where $\de$ is the covariant derivative with respect to
the round metric on $S^3$. In addition, because of eq. ($A.6$)
of the Appendix, the $\rho^i$ obey the symmetry relations,
$$
\r^i_{_{\mo\mt}} = -\tilde \r^i_{_{\mo\mt}} \equiv - \e_{_{M_1}}\e_{_{M_2}}
\r^i_{_{-M_2\,-M_1}} \ .
\equn\put\symrel
$$
These relations together with {\rrel} reduce the number of linearly
independent real
vectors to precisely the required six Killing vectors of $S^3$.

The constant Killing vector {\time} which generates time translations on
$R \times S^3$ and is conformally mapped to the global dilation generator
of flat
Minkowski spacetime is clearly a solution of eq. {\ckv} as well.

The remaining eight proper conformal Killing vectors, denoted by
$\k^{a(\pm)}_{_M}$ are easily found in terms of the spherical harmonics on
$S^3$
in the form,
$$
\k^{(+)}_{_M} = -\hbox{${\sqrt V} \over 2$}e^{it}\,(\Yh^{\ast} ,\, i \vec\de\,
\Yh^{\ast})=  \left[\k^{(-)}_{_M}\right]^{\ast}\ ,
\equn\put\pcv
$$
These eight $\k^{(\pm)}_M$ correspond in flat space to four translations
plus four special conformal transformations, and satisfy
$$
\de_a \k^{a(\pm)}_{_M} = \dot \k^{0(\pm)}_{_M} + \de_i\, \k^{i(\pm)}_{_M}
 = \pm 4 i \k^{0(\pm)}_{_M}\ .\equn\put\divpcv
$$

Together these $6+1+8=15$ conformal Killing fields generate the Lie algebra
of the $SO(4,2)$ group of conformally flat $4$ dimensional spacetime.
Classically this may be seen at the level of the algebra of the Lie
derivatives. Recall that the action of an infinitesimal coordinate
transformation generated by an arbitrary vector $\x^{a}$ on a field density
$\f$ is
$$
\d_{\x} \f = \x^{a} \de_{a} \f +\hbox {$w \over 4$} \f\, (\de_{a}\x^{a})\ ,
\equn\put\cwei
$$
where $w$ is, by definition, the conformal weight of $\f$. From this definition
we find that the effect of two infinitesimal transformations $\x$ and
$\z$ performed in different orders is
$$
\left[ \d_{\x}, \d_{\z} \right] \f = \d_{[\x , \z]} \f ,\equn
$$
where $[\x,\z]$ is the classical Lie bracket defined by {\lbrdef}.

If we now use these general properties of Lie derivatives applied to the $15$
conformal Killing fields of $R \times S^3$, we find
$$\eqalign{
\left[\h,\k^{(\pm)}_{_M}\right] &= \mp \k^{(\pm)}_{_M}\cr
\left[\k^{(\pm)}_{_{M_1}}, \k^{(\mp)}_{_{M_2}}\right] &= \pm 2\h \d_{_{M_1M_2}}
+ 2 \r_{_{M_1M_2}} \cr
\left[\k^{(\pm)}_{_M}, \rho_{_{M_1M_2}}\right] &= \d_{_{MM_2}}
\k^{(\pm)}_{_{M_1}} - \e_{_{M_1}}\e_{_{M_2}}\d_{_{M\,-M_1}}
\k^{(\pm)}_{_{-M_2}} \cr
[\rho_{_{M_1M_2}}, \rho_{_{M_3M_4}}] &= \d_{_{M_1M_3}}\rho_{_{M_2M_4}} + {\rm
permutations}\ ,\cr}
\equn\put\liebr
$$
with all other Lie brackets vanishing or obtained from these by symmetry
relations. This classical algebra of Lie brackets is isomorphic to $SO(4,2)$
and will go over to the quantum commutator Lie algebra with the important
modification of a c-number shift in the right side of the second relation in
{\liebr}, as we shall show in the next section.

The quantum generators corresponding to these $15$ conformal Killing
vectors may now be constructed. The Hamiltonian has been given already by
{\ham}.
The generators of rotations on $S^3$ are
$$
R_{_{\mo\mt}} \equiv  X_{\r_{_{M_1M_2}}} = -\int_{S^3}\, d\O\
\r_{_{i\,M_1M_2}}\,
T^{0i}\ ,
\equn\put\genR
$$
which are time-independent volume preserving diffeomorphisms. Since the
physical
states must be invariant under these particular volume preserving spatial
diffeomorphisms, we require
$$
R_{_{\mo\mt}}\vert phys\rangle = 0\ . \equn
$$
The generators of the eight proper conformal Killing transformations are
$$\eqalign{
K^{(\pm)}_{_M}  &\equiv -\int_{S^3} \, d\O\, \k^{(\pm)}_{a\,_M}\, T^{0a}\cr
& = -\int_{S^3}\, d\O\, \k^{(\pm)}_{0\,_M} \left(T^{00} \pm i
\pa_t T^{00}\right)\cr
& = \mp 2i\int_{S^3}\,d\O\, \k^{(\pm)}_{0\,_M}\, \pa_t T^{00(\pm)}\cr
& = e^{\pm it}\, \dot T^{(\pm)}_{{1\over 2}M}\cr }
\equn\put\genK
$$
where we have used {\pcv}, the conservation of $T^{ab}$ an integration by
parts,
and the following definition for the moments of the time derivative of the
energy
density:
$$
\dot T^{(+)}_{_{JM}} \equiv -i\sqrt{V} \int_{S^3} d\O \, Y^*_{_{JM}}\,
\pa_tT^{00(+)}\
.\equn\put\enmom
$$
In passing to the third line of {\genK} we have also made use of the fact that
the Noether charges $K^{(\pm)}$ are time-independent, so that the exponential
time dependence in $\k^{(\pm)}_{0\,_M}$ must be canceled by that in $T^{00}$.
In other words, the $\k^{(\pm)}_{0\,M}$ act as projectors onto
time dependence $\exp(\mp it)$ of $\pa_tT^{00}$ which selects precisely the
positive or negative frequency moments, $\dot T^{(\pm)}_{{1\over 2}M}$ of
the last line, provided the energy-momentum tensor is traceless.
Hence the positive frequency component, $K_M^{(+)}$ is just proportional to
$\dT$ for $J={1\over 2}$, and is automatically enforced to vanish on the
physical
states from our general condition {\Xphys}.

The above discussion shows that $8+6=14$ of the global special conformal
transformations of $R \times S^3$ are identified with the lowest moments of
spatial
diffeomorphism generators $X_{\x}$, once they are separated into positive,
negative
and zero frequency components. Together with the Hamiltonian $H$ they form
the Lie
algebra of the global conformal group $SO(4,2)$. In two dimensions the
conformal
group is infinite dimensional, and this identification of diffeomorphism
generators
with conformal group generators is extended to all higher moments. There are an
infinite number of conserved currents and charges of the conformal Virasoro
algebra,
and as shown in {\Ttwo} they are in one-to-one correspondence with the
generators
of positive, negative and zero frequency spatial diffeomorphism generators,
together with the Hamiltonian $L_0 + \bar L_0$. This extended Virasoro
algebra is
also essential to the understanding of the invariance of the theory under
finite
diffeomorphisms.  As we see by the above discussion, this situation does not
generalize to higher dimensions, where the conformal group is finite
dimensional.
This means, in particular that the commutators of the higher moments of the
separate positive and negative frequency components of the spatial
diffeomorphism
generators $X_{\x}^{(\pm)}$ do not necessarily form a closed algebra, or if
they
do, this algebra is unknown. Hence the full algebraic structure of
diffeomorphism
invariance in four dimensions is not determined by our approach, and remains an
open problem.

\vfill
\eject

\newcount\eqncount
\sectadv
\set\sect{3}
\beginsection{3. The Quantum Hamiltonian Constraint}

In this section we wish to calculate the quantum contribution to the
Hamiltonian constraint on the product space $R \times S^3$. The
Hamiltonian physical state condition will be of the form,
$$
H|phys> = a|phys>
$$
where $a$ is the finite subtraction constant we wish to determine. The
classical value of $a=0$ will turn out to be untenable at the quantum level.
As in string theory on the two dimensional world sheet, this
subtraction may be determined by consistency with the global conformal algebra
of the Fadeev-Popov ghost-antighost system. Before entering into the
detailed analysis of ghosts, let us give a heuristic argument why
a value different from zero should be expected, based on experience in $D=2$.

Any physical state in a diffeomorphic invariant theory should be created by
an invariant scalar, which may be expressed as the spacetime integral
$\int d^4 x$ of an operator with conformal weight $4$ ({\it i.e.} a volume
density), operating  on the vacuum state invariant under the symmetry group
of $R \times S^3$, $|0\rangle$,
$$\
\vert phys \rangle = \co_4 \vert 0\rangle\ ,
\equn\put\creat
$$
which is annihilated by $H$,
$$
H \vert 0\rangle = 0\ .\equn\put\Hkill
$$
Now under the conformal transformation which maps $R \times S^3$ to
flat Euclidean space {\cmap} the Hamiltonian generator of time translations
is mapped to the global dilation generator of flat space, which we denote
by $\bar
H$. In particular this means that in flat space
$$
\left[\bar H,\bar \co_4\right] = 4\bar\co_4\ ,
\equn\put\soinv
$$
with $\xi^a = x^a$ and $w=4$ in eq. {\cwei}.
Performing the transformation of this relation back to $R\times S^3$ allows
us to remove the overbars. Then operating on the invariant vacuum and
using {\creat} and {\Hkill} yields
$$
H \vert phys\rangle = 4 \vert phys \rangle\ .
\equn\put\four
$$
We conclude on the basis of these simple dimensional considerations that $a=4$.

This interpretation is completely consistent with the corresponding value of
$H$ in closed string theory, namely $2$ on the world-sheet of $2$ spacetime
dimensions. At the canonical algebra level this $2$ arises from the anomalous
commutator in the global conformal algebra $SU_L(1,1) \times SU_R(1,1) \cong
SO(2,2)$ of the ghost energy-momentum tensor, introduced in order to fix
the gauge
completely in path integral quantization of the string. There is no anomaly
in the
global algebra of matter energy-momentum tensors, so that the ghosts alone
contribute to the shift of $a$ to $2$ from its classical value of $0$. In order
to
justify the heuristic dimensional argument leading to eq. {\four} we shall
need to
carry out the analogous construction of the ghost energy-momentum tensor and
the
global algebra of its moments in four dimensions.

Let us first recall the quantum measure in the path integral over metrics in
the conformal decomposition {\confdef}. In several earlier papers
[{\cFIG}-{\cAMM}]
we introduced the natural orthogonal splitting on the space of metric
deformations
induced by this decomposition, namely,
$$
{\d}g_{ab}=\left(2{\d\s}+ \hbox{$1 \over 2$}{\de}_c{\x}^c\right)
g_{ab}+{\left(L\xi\right)}_{ab}+ h^{\perp}_{ab} ,
\equn\put\defo
$$
where $h^{\perp}_{ab}$ is a transverse, tracefree symmetric tensor, and $L$ is
defined by {\ckv}. Then the path
integral measure on the space of all metrics may be written as a product over
the three coordinates $(\s , \x , h^{\perp})$ together with a Jacobian for
the change of coordinates, given by
$$
J=\left[{\rm det}'\left(L^{\dag}L\right)\right]^{1 \over 2}\ ,
\equn\put\jac
$$
where the prime indicates the restriction to the non-zero mode subspace of the
operator $L$ defined in {\ckv}. The two operators $L$ and $L^{\dag}$ are
individually conformally covariant,
$$
L = e^{2 \s} \bar L e^{-2 \s}\qquad , \qquad
L^{\dag} = e^{-4 \s} \bar L^{\dag} e^{2 \s}, \equn\put\Ltrans
$$
although their product $L^{\dag}L$ is not.
Let us remark that since $L$ maps vectors into symmetric, traceless tensors
and $L^{\dag}$ maps traceless, symmetric tensors into vectors, there
is no meaning to the determinants of each of these operators individually.
However, because of the one-to-one correspondence between eigenvectors of
$L^{\dag}L$, $\x^{(\l)}$ with eigenvalue $\l$ and eigentensors of
$LL^{\dag}$ in the
range of $L$ by the relation,
$$
(LL^{\dag})(L\xi^{(\lambda)}) = L(L^{\dag}L)\xi^{(\lambda)} = \lambda
(L\xi^{(\lambda)}), \equn\put\eigen
$$
the spectrum of $L^{\dag}L$ is the same as that of $LL^{\dag}$, and
the determinants of these two operators are equal when
restricted to their non-zero mode subspaces. Hence the Jacobian {\jac}
may be reproduced by introducing the ghost action,
$$
S_{gh} = -{i\over 4\p^2}\int d^4x \sqrt{-g}\ b^{\a\b}(Lc)_{\a\b}\ ,
\equn\put\Sgh
$$
where all fields and operators now transform covariantly under the
conformal decomposition {\confdef}.
The integration over the anticommuting Grassman fields $b^{\a\b}$
and $c_{\a}$, restricted to the non-zero mode subspaces of $L$ and
$L^{\dag}$, yields precisely the Jacobian $J$ of eq. {\jac}. The normalization
factor in $S_{gh}$ is arbitrary at this point, and is chosen in order to yield
a conveniently normalized hermitian energy-momentum tensor for the ghosts
below.

One complication to this ghost construction arises from the zero modes of
$L$ and
$L^{\dag}$. The zero modes of $L$ in {\ckv} are the conformal Killing
vectors $\o$
of the background corresponding to the generators of the finite dimensional
global
conformal group $SO(4,2)$ in the conformally flat $R \times S^3$ metric.
The zero
mode space of $L^{\dag}$ is spanned  by all transverse, tracefree symmetric
tensors
which is an infinite dimensional space for $D=4$. The reason for the difference
with $D=2$ is clear: there are no physical transverse graviton modes in
$D=2$ where
all metrics are locally conformally flat. It is only by including the
contributions
of the zero mode spaces of $L$ and
$L^{\dag}$ that the $\s$ dependence of measure may be extracted in the form of
a local action
$\G[{\bar g};\s]$ and the gravitational measure presented in the
factorized form,
$$
[\cd(e^{2\s}{\bar g})]= e^{-\G[{\bar g};\s]}
\left({{\rm det}'L^{\dagger}L\over {\rm det}<{\o}_j|{\o}_k>}
\right)^{1/2}_{\bar g}[\cd{\it {Vol}}(Diff_0)]
[\cd{\s}][\cd{\bar g}^{\perp}]\ .
\equn\put\measure
$$
In this form the integration over the gauge orbit of coordinate
diffeomorphisms {\it Vol(Diff}$_0$) (continuously connected to the identity)
has
been factorized explicitly and may be divided out. The $\s$ dependence of the
measure is expressed purely by the local action $\G[{\bar g};\s]$ which was
obtained in refs. [\cAntMot] and [\cAMM], and is studied in detail in paper
II. The
physical graviton modes  are contained in $\bar g^{\perp}$ which must be
integrated
with the measure $\cd{\bar g}^{\perp}$ together with some invariant local
action
$S_{inv}[{\bar g}]$ in $D=4$.

The existence of a finite number of conformal Killing vectors $N_{CKV}=15$
means
that the dimension of the range of $L$ has decreased by $N_{CKV}$  from the
generic
non-conformally flat background metric. Since the dimension of the space of
metrics
should not decrease discontinuously by increasing the symmetry of the
background
through continuous deformations, these $N_{CKV}$ diffeomorphisms should
reappear in
the other parts of the general decomposition of metric deformations of {\defo}.
Phrased differently, when there are $N_{CKV}$ zero modes of $L$ the
condition that
the traceless part of $\d g_{ab}$ satisfy the Lorentz-Landau gauge condition
$$
\nabla^{b}h^{\perp}_{ab} = 0
\equn\put\gaugecond
$$
does not fix the gauge completely, and there should be precisely $N_{CKV}$
additional transverse, tracefree tensors $h^{\perp}_{ab}$ (or scalars $\s$)
corresponding to this residual gauge freedom.  It is not difficult to
construct the
transverse, traceless, symmetric tensor corresponding to each CKV of $R
\times S^3$. To do so let us present the general $h^{\perp}_{ab}$ in the
form,
$$
h^{\perp}_{ab} = \left(\matrix{u&v_i\cr v_i&{u\over
3}g_{ij} + h_{ij}^{TT}\cr}\right),
\equn\put\tt
$$
where the $h_{ij}^{TT}$ are transverse and traceless with respect to the
spatial metric on $S^3$, and
$$\eqalign{
\dot u &= \de_i v^i \cr
\dot v_i &= \hbox{$1 \over 3$} \de_i u \cr}
\equn\put\transv
$$
are the conditions that $h_{ab}$ be transverse with respect to the
full four-metric on $R \times S^3$. Now the key observation in the
construction is to recognize the similarity between the transversality
conditions {\transv} and the time-time and time-space components of
the conformal Killing equations {\ckv} $L\o = 0$, namely,
$$\eqalign{
\dot \o^0 &= \hbox{$1 \over 3$} \de_i \o^i \cr
\dot \o_i &= \de_i \o^0\ . \cr}
\equn\put\ckvs
$$
By identifying
$$\eqalign{
u &\rightarrow  3\o^0 , \qquad {\rm and} \cr
v_i &\rightarrow  \o_i \cr}
\equn\put\identif
$$
we observe that any solution of {\ckvs} generates a transverse, tracefree
tensor
{\tt}. Of the infinite number of transverse,
tracefree tensors satisfying {\transv} we thereby select only those satisfying
also the space-space part of $L\o = 0$, namely
$$
\nabla_i \o_j + \nabla_j \o_i = \hbox{$1\over 2$}g_{ij} (\dot \o^0 + \de_k\o^k)
= \hbox{$2 \over 3$} g_{ij} \nabla_k\o^k\ .
\equn\put\ckvsp
$$
and no $h_{ij}^{TT}$ physical graviton component.

Since $R\times S^3$ has exactly $N_{CKV} = 15$ solutions of the equations
{\ckvs} and {\ckvsp}, the above identification constructs exactly $15$
transverse, tracefree tensors corresponding to the conformal Killing vectors.
The fact that these transverse tensors are in one-to-one correspondence with
conformal Killing vectors and independent of the detailed dynamics of the
physical graviton modes implies that they should not appear in the
expansion of any generally covariant action around a background possessing
these conformal Killing symmetries. Indeed, we have checked explicitly that
for the Weyl tensor squared action, when expanded
to quadratic order around $R \times S^3$ these $15$ tensor modes drop out of
the
quadratic action, {\it i.e.} are solutions of the differential equation
obtained
from the action. These $15$ solutions are the only modes which are not in the
ordinary physical space of solutions of the fourth order wave equation of the
linearized Weyl theory.

In the case of the Einstein-Hilbert action the $R \times S^3$ geometry is
not a solution of the equations of motion in the absence of matter. However,
if a positive cosmological term is added to the action then de Sitter spacetime
is a conformally flat solution with the global topology of $R \times S^3$,
possessing exactly $15$ conformal Killing vectors. In the expansion about this
background one finds once again exactly $15$ modes ($10$ in $h^{\perp}$ and
$5$ in
$\s$ in this case) which drop out the quadratic action, and which are not
contained
in the ordinary
$h_{ij}^{TT}$ physical graviton solutions of the wave equation.

With this understanding of the origin of the ghost action {\Sgh} in the
covariant path integral and the special role played by the zero modes
of $L$ we turn now to the canonical quantization procedure on the space
$R \times S^3$. The ghost system must be quantized together with the
$\s$, $h^{\perp}$ and matter fields of the theory. The equations of
motion for the ghost field $c_{\a}$ obtained by varying $S_{gh}$ with
respect to $b^{\a\b}$ is
$$
(L c)_{\a\b} = 0\ ,\equn
$$
so that $c_{\a}$ may be expanded in terms of the $15$ conformal Killing
vectors of $R \times S^3$,
$$
c^{\a}(x) = \h^{\a}\, c_0 + \rho^{\a}_{_{M_1M_2}}(x)\, c_{_{M_1M_2}} +
\k_{_M}^{{\a}(-)}(x)\,c_{_M} + \k_{_M}^{{\a}(+)}(x)\,c_{_M}^{\dag}\ .
\equn\put\vecm
$$
The tensor ghost field $b_{\a\b}$ may be expanded then in the corresponding
$15$ transverse, tracefree modes constructed by relations {\tt} and
{\identif} above:
$$\eqalign{
b_{00}(x) &= u_0\, b_0 + u_{_M}(x)\,b_{_M} + u^*_{_M}(x)\,b^{\dag}_{_M}\cr
b_{0i}(x) &= v_{i\ _{M_1 M_2}}(x)\, b_{_{M_1M_2}} + v_{i\ _M}(x)\,b_{_M} +
v^*_{i\ _M}(x)\, b^{\dag}_{_M}\cr
b_{ij}(x) &= \hbox{$1 \over 3$} g_{ij}\left(u_0\, b_0 + u_{_M}(x)\,b_{_M} +
u^*_{_M}(x)\,b^{\dag}_{_M}\right)\ . \cr }
\equn\put\tenm
$$
The normalization of the tensor modes may be fixed by taking
$$\eqalign{
u_0 &= \h^0=1\cr
u_{M} &= 2\k^{0(-)}_{M}\cr
v^i_{M} &= \hbox {$2 \over 3$} \k^{i(-)}_{M}\cr
v^i_{\ M_1M_2} &= \hbox {$1\over 2$} \rho^{i\ *}_{M_1M_2}\cr}
\equn\put\tenvec
$$
so that the variation of the ghost action takes the canonical form, and
canonical anti-commutation relations may be imposed:
$$\eqalign{
\{b_0,c_0\} &=1 \cr
\{b_{_{M_1}},c^{\dag}_{_{M_2}}\} &=
\{b^{\dag}_{_{M_1}},c_{_{M_2}}\} = \d_{_{M_1M_2}}\cr
\{b_{_{M_1M_2}}, c_{_{M_3M_4}}\} &= \d_{_{M_1M_3}}\d_{_{M_2M_4}}
 - \e_{_{M_1}}\e_{_{M_2}} \d_{_{M_1\,-M_4}}\d_{_{M_2\, -M_3}}\cr}
\equn\put\gcg
$$
where $\e_{_{M}}$ is defined in ($A.6$).

The energy-momentum tensor of the ghosts obtained by varying $S_{gh}$ is
$$
T^{\m\n}_{gh}= {i\over 2\p^2}\left(\de_{\a}(b^{\m\n}c^{\a}) + 2
b_{\a}^{(\m}\de^{\n)}c^{\a} - {1\over 2}b^{\m\n}\de_{\a}c^{\a}\right)\ .
\equn\put\ghener
$$
By substituting the mode expansions {\vecm} and {\tenm} into this expression
and using {\tenvec}, {\ks} and {\pcv}, we obtain
$$
H^{gh} \equiv \int_{S^3}\, d\O\, :T^{00}_{gh}: = \sum_M
\left(b^{\dag}_{_M} c_{_M} + c^{\dag}_{_M} b_{_M}\right)\ ,\equn
$$
which is the Hamiltonian of the finite dimensional ghost system in normal
ordered form. The other non-vanishing moments of the ghost energy-momentum
tensor we require are
$$\eqalign{
K^{(+)\, gh}_{_M} &\equiv  -\int_{S^3}\, d\O\,
\k^{(+)}_{a\,_{MM'}}\,T^{0a}_{gh}
 \cr &=2b_0 c_{_M} + b_{_M}c_0 + \sum_{_{M'}}\left(
2b_{_{M'}}c_{_{M'M}} + b_{_{MM'}}c_{_{M'}}\right)\cr
K^{(-)\, gh}_{_M} &\equiv  -\int_{S^3}\, d\O\,
\k^{(-)}_{a\,_M}\,T^{0a}_{gh}\cr &=  2 c^{\dag}_{_M} b_0 + c_0
b^{\dag}_{_M} + \sum_{M'}\left(2c_{_{MM'}}b^{\dag}_{_{M'}} +
c^{\dag}_{_{M'}}b_{_{M'M}}\right)\cr}\equn
$$
and
$$\eqalign{
&R^{gh}_{_{M_1M_2}} \equiv  -\hbox {$1\over 2$}\int_{S^3}\, d\O\,
\r_{i\,_{M_1M_2}}\,\left(T^{0i}_{gh}[b,c] - T^{0i}_{gh}[c,b]\right) =\cr
&\sum_M \left(b_{_{MM_2}}c_{_{MM_1}} - b_{_{M_1M}}c_{_{M_2M}}\right)
+ b^{\dag}_{_{M_2}}c_{_{M_1}} - b_{_{M_1}}c^{\dag}_{_{M_2}} -
s_{_{M_1}}s_{_{M_2}}
\left(b^{\dag}_{_{-M_1}}c_{_{-M_2}} - b_{_{-M_2}}c^{\dag}_{_{-M_1}}\right)\cr
&\qquad - \left(b\leftrightarrow c\right)\ ,\cr}\equn
$$
where this last expression has been anti-symmetrized in the ghost and
anti-ghost operators.

With these expressions for the generators of the global conformal group,
and making use of the properties of the first few scalar and vector harmonics
catalogued in the Appendix, it is a straightforward (but tedious) exercise to
compute their commutators and verify that they do close to the algebra of
$SO(4,2)$
with the important modification of a {\it c-number shift} of $-4$ in the
Hamiltonian $H^{gh}$. This shift may be found by calculating the commutator,
$$
\left[ K^{(+)\, gh}_{_{M_1}}\,, K^{(-)\,gh}_{_{M_2}}\right]
= 2\d_{_{M_1M_2}}\left(H^{gh} - 4 \right) +
2R^{gh}_{_{M_1M_2}}
\equn\put\shift
$$
The most rapid way to obtain the shift of $-4$ is to calculate the
expectation value of the above commutator in the ghost vacuum
state which is annihilated by the positive frequency destruction operators,
$c_{_M}$ and $b_{_M}$ while for the zero frequency operators it
may be chosen to obey (see eqs {\gcg}):
$$\eqalign{
c_0\vert 0\rangle &= c_{_{M_1M_2}}\vert 0\rangle = 0 \qquad {\rm but}\cr
\langle 0 \vert b_0 &= \langle 0 \vert b_{_{M_1M_2}} =0\ .\cr} \equn
$$
This state is invariant under the maximal subgroup of $SO(4,2)$ generated by
$H^{gh}, K^{(+)gh}_{_M}$ and $R^{gh}_{_{M_1M_2}}$. Then we find
$$\eqalign{
\langle 0 \vert \left[ K^{(+)\, gh}_{_{M_1}}\,, K^{(-)\,gh}_{_{M_2}}\right]
\vert
0\rangle &= 2 \langle 0 \vert b_{_{M_1}}c_0 c^{\dag}_{_{M_2}}b_0 \vert 0
\rangle +
2 \sum_{_{MM'}}\langle 0\vert
b_{_{M_1M}}c_{_M}c_{_{M_2M'}}b^{\dag}_{_{M'}}\vert0\rangle\cr &=
-2\d_{_{M_1M_2}}
-2 \sum_{_{M'}} \left( \d_{_{M_1M_2}}\d_{_{M'M'}} -\d_{_{M_1M'}}\d_{_{M'M_2}}
\right) \cr
&= -8 \d_{_{M_1M_2}} \cr} \equn
$$
which is equivalent to a shift of $-4$ in $H^{gh}$ from {\shift}.

The shift of $-4$ is completely quantum mechanical in origin and comes
only from the $bc$ ghost system. Matter fields
exhibit no such shift, as we shall see explicitly in the example
of the next section and in paper II.
As a consequence, the Hamiltonian constraint which is the
lowest moment of the classical Wheeler-DeWitt condition $T^{00} = 0$
is {\it modified by a definite c-number shift}, and
obeys {\four} on any physical state of a coordinate invariant theory
with zero energy-momentum trace $T_a^a = 0$ on $R\times S^3$.

\vfill
\eject

\newcount\eqncount
\sectadv
\set\sect{4}
\beginsection{4. Quantization of a Conformal Scalar on $R \times S^3$}

As a concrete realization of the general structure of the diffeomorphism
generators on $R \times S^3$ we present here the results for the Fock
representation in the simplest example, a conformal free scalar field
with action,
$$
S = -\hbox{$1 \over 2$} \int d^4x\sqrt{-g}\, \left(
g^{ab}\nabla_a\Phi\nabla_b\Phi +
\hbox{$1 \over 6$} R \Phi^2\right)\ .\equn
$$
On $R \times S^3$ with unit radius, the field $\Phi$ may be
expanded in functions of the form $\exp (-i \o t) \Y$, and
the wave operator $\sq - 1$ factorizes as
$$
(\sq - 1)\exp (-i \o t) \Y = \left(\o^2- (2J+1)^2\right)\exp (-i \o t) \Y\
,\equn
$$
so that $\Phi$ may be written in terms of creation and destruction
operators
$$
\Phi = \sum_{JM}{1\over \sqrt{2(2J+1)}}\left( e^{-i(2J + 1)t}\,\Y\ph
+ e^{i(2J + 1)t}\,\Ys\phd\right)\ ,\equn\put\fock
$$
obeying canonical commutation relations
$$
\left[\varphi_{_{J_1M_1}}, \varphi^{\dag}_{_{J_2M_2}}\right] =
\d_{_{J_1J_2}}\d_{_{M_1M_2}}\ .\equn
$$
The (normal ordered) energy-momentum tensor for the scalar field is
$$
T^{ab} = :\hbox{$2 \over 3$} \nabla^a\Phi\nabla^b\Phi - \hbox{$1 \over
3$}\Phi\nabla^a\nabla^b\Phi -\hbox{$1 \over 6$}g^{ab}\left((\nabla\Phi)^2 +
\hbox{$1 \over 6$}\Phi^2\right) + \hbox{$1 \over 6$}R^{ab}\Phi^2:
\ ,\equn\put\enermom
$$
which is classically traceless. Furthermore, all three
curvature invariants appearing in the trace anomaly, $C_{abcd}C^{abcd}, \sq R$
and $G = \Rie - 4 \Ric + \R $ vanish on $R\times S^3$, so that the trace of the
energy-momentum tensor {\enermom} also vanishes in the quantum theory:
$$
T^a_a = 0 \qquad {\rm on}\ \ R \times S^3.
\equn
$$

On $S^3$ with unit radius, $R_{0a}=0, R_{ij} = 2 \d_{ij}$ and the energy
density
becomes
$$
T^{00}\big\vert_{_{R\times S^3}} = :\hbox{$1 \over 2$} \dot\Phi^2 -\hbox{$1
\over 2$} \Phi\ddot\Phi + \hbox{$1 \over 12$} \nabla^2(\Phi^2):\ ,
\equn\put\enphi
$$
while
$$
T^{0i}\big\vert_{_{R\times S^3}} = :-\dot\Phi \nabla^i\Phi +
\hbox{$1 \over 6$}\nabla^i\pa_t(\Phi^2):\ .\equn
$$
Substituting the mode expansion {\fock} into {\enphi} we form the moments,
$$\eqalign{
\dT\Big\vert_{t=0} &= -\sum_{J_1M_1\atop J_2M_2}
\cc^{JM}_{\jo\mo\jt\mt}\Biggl\{(\jo
+ \jt +1) \left[(\jo-\jt)^2 - \hbox{$1 \over 3$}J (J+1)\right]
\varphi_{_{J_1M_1}}
\varphi_{_{J_2M_2}}\cr +&(\jo-\jt)\left[(\jo +\jt + 1)^2 - {1\over 3}J
(J+1)\right]
\tilde\varphi^{\dag}_{_{J_2M_2}} \varphi_{_{J_1M_1}}\Big\vert_{_{\jo
>\jt}}\Biggr\}\cr}
\equn\put\scdT
$$
as defined by {\enmom}. Here the notation
$$
\tilde\ph \equiv \e_{_M} \varphi_{_{J\ -M}} \equn
$$
has been used, and the symbol $\cc$ is an $O(4)$ angular momentum coupling
coefficient defined by
$$\eqalign{
\cc^{JM}_{\jo\mo\jt\mt} &\equiv  {\sqrt {V}\over \sqrt{(2\jo+1)(2\jt + 1)}}
\int_{S^3} d\O\  \Ys Y_{_{\jo\mo}} Y_{_{\jt\mt}}\cr &= {1\over \sqrt{(2J + 1)}}
C^{Jm}_{\jo m_1\jt m_2}C^{Jm'}_{\jo m'_1\jt m'_2}\ ,\cr}
\equn\put\cgpro
$$
where eq. ($A.7$) of the Appendix has been used and $C$ denotes an
ordinary $SU(2)$ Clebsch-Gordon coefficient.
In the case of the lowest non-trivial moment $J= {1\over 2}$ we obtain the
realization of the conformal generator $K^{(+)}_{_M}$ in {\genK} for the
conformal
scalar field,
$$
K^{(+)}_{_M} = -\sum_{JM_1M_2} \cc^{\ha M}_{J+\ha\mo J\mt}(J+1)(2J+1)
\tilde\varphi^{\dag}_{_{JM_2}} \varphi_{_{(J+\ha) M_1}}\ .
\equn\put\Kscl
$$

As discussed previously the moments $\dT + {\dot{T}}_{JM}^{(-)}$ generate
the volume
non-preserving spatial diffeomorphisms on $S^3$. The volume preserving
diffeomorphisms are generated by $X_{\cy^*_{J\cm}}\equiv X_{_{J\cm}}^{(+)} +
X_{_{J\cm}}^{(-)}$ with
$$
X_{_{J\cm}}^{(+)}\Big\vert_{_{t=0}} =
-i\sum_{\jo\mo\atop\jt\mt}\ck^{J\cm}_{\jo\mo\jt\mt}\left\{ (\jo-\jt)
\varphi_{_{\jo\mo}}\varphi_{_{\jt\mt}} + (\jo +\jt + 1)
\tilde\varphi^{\dag}_{_{J_2M_2}}\varphi_{_{J_1M_1}}\right\}_{_{\jo \ge \jt}}
\equn\put\scX
$$
where the coupling coefficient with the transverse vector $O(4)$ spherical
harmonic
$\cy^i_{J\cm}$ is defined by
$$
\ck^{J\cm}_{\jo\mo\jt\mt} \equiv {1\over 2\sqrt{(2\jo+1)(2\jt + 1)}}
\int_{S^3} d\O\  \cy^{i*}_{J\cm}\, Y_{_{\jo\mo}}\nabla_i Y_{_{\jt\mt}}\ .
\equn\put\vcoup
$$
Relevant properties of the transverse vector harmonics on $S^3$ are
given in the Appendix. The lowest moment of the volume preserving
diffeomorphisms
corresponding to $\x^i = \r^i$ of eq. {\ks} are precisely
the $6$ rotation generators of $SO(4)$ for the conformal scalar field,
$$
R_{_{\mo\mt}} = {i\over 2} {\sum_{JM_3M_4}} \int_{S^3} d\O\,
\r^i_{_{\mo\mt}} Y_{_{JM_4}}^* \nabla_i Y_{_{JM_3}}
\varphi^{\dag}_{_{J M_4}} \varphi_{_{J M_3}}\ .
\equn\put\Rscl
$$

Finally the Hamiltonian generator for the conformal scalar field on
$R \times S^3$ is simply
$$
H = \sum_{JM} (2J + 1) \phd\ph\ . \equn\put\Hscl
$$
With the generators of the $SO(4,2)$ global conformal algebra for
the scalar field {\Kscl}, {\Rscl} and {\Hscl} in hand, one may
verify that the quantum generators
obey the algebra corresponding to the {\it classical} Lie brackets
in {\liebr}, using the properties of the harmonics listed in the Appendix.
There is {\it no} shift in the Hamiltonian for the
matter field, in contrast to the $-4$ found in the $bc$ ghost
calculation of the previous section, since there is no analog of
the ghost operator ordering question in the scalar sector.

We remark that the classical algebra of spatial diffeomorphisms
as given by {\spat} is respected by the sum of positive
and negative frequency generators constructed from the scalar
field Fock operators above. However, taking commutators of the higher
moments with {\it separate} positive and negative frequency parts of
the diffeomorphism generators yields complicated expressions
with no clear algebraic structure apparent, in contrast to the case in
two spacetime dimensions where the full Virasoro algebra emerges this way.
As mentioned in Section $2$, whether or not such an extended algebra
should exist, and if it does not what the consequences would be for
the consistency of reparametrization invariant quantum theories
in four spacetime dimensions remains an interesting open problem.

In order to illustrate the method for constructing physical states
which are invariant under infinitesimal coordinate diffeomorphisms, let us now
apply the constraints to the general states of the matter field
in the absence of gravitational field modes.
The matter states are constructed in the usual way by operating with any
number of creation operators $\phd$ on the Fock vacuum obeying
$$
\ph \vert 0\rangle = 0\ , \equn
$$
on which $H$, $K^{(+)}_{_M}$ and $R_{_{\mo\mt}}$ also vanish. In analogy
with the canonical method in two dimensions we define the level of any
eigenstate of the Hamiltonian by its integer eigenvalue,
$$
H \vert N\rangle = N \vert N\rangle \ . \equn
$$
{}From the expression for the scalar field Hamiltonian {\Hscl} we observe
that each creation operator $\phd$ contributes $2J + 1$ units to the
level of the state. Hence we may find the physical states of the scalar field
by writing first the general linear combination of
creation operators that create a given level $N$ state, and then
applying the diffeomorphism constraints to relate the coefficients
of the linear combination of coefficients at this level, to see which states
survive. This procedure is best carried out by beginning with the lowest
levels and moving up in $N$ one unit at a time.

At level one the only possible state is
$$
\vert 1\rangle = \varphi^{\dag}_{00}\vert 0\rangle \equn
$$
which survives imposition of all the spatial diffeomorphism constraints
$$
\dT \vert 1\rangle = X_{_{J\cm}}^{(+)}\vert 1\rangle = 0\ . \equn
$$

At level two the general state is of the form
$$
\vert 2\rangle = \left(\a (\varphi^{\dag}_{00})^2 + \b_M
\varphi^{\dag}_{{1\over
2}M}
\right) \vert 0\rangle \ .\equn
$$
Requiring that $\dot T_{{1\over 2}M}$ on this state vanishes yields $\b_M =0$,
whereas the first term in the combination survives all the constraints for
general $\a$.

At level three the general state is
$$
\vert 3\rangle = \left(\a^{\prime} (\varphi^{\dag}_{00})^3 + \b_M^{\prime}
\varphi^{\dag}_{00}\varphi^{\dag}_{{1\over 2}M} + \g_M^{\prime}
\varphi^{\dag}_{1M} \right) \vert 0\rangle \ .\equn
$$
Now the application of the $\dot T_{{1\over 2}M}$ and $\dot T_{1M}$ constraints
require $\b_M^{\prime} = \g_M^{\prime} = 0$ while the first term again survives
all the constraints. Indeed applying the higher $J$ constraint first eliminates
the $\g_M^{\prime}$ term and then the $\b_M^{\prime}$ term is eliminated
for the same reason as at level $2$.

Continuing in this way it is not difficult to convince oneself inductively
that the only state at level $N$ which survives all the spatial diffeomorphism
constraints is
$$
\vert N\rangle = (\varphi^{\dag}_{00})^N \vert 0\rangle \ .\equn\put\scN
$$
That this state does indeed satisfy all of the spatial constraints is easy
to see directly from the forms of {\scdT} and {\scX}. For example, in the
first term
of $\dT$ the only non-trivial terms result from commuting the destruction
operators with $\jo = \jt =0$ through the creation operators in {\scN} above.
But then the coupling coefficient $\cc$ is non-vanishing only for $J=0$
which is multiplied by the coefficient $[(\jo-\jt)^2 - {1\over 3}J (J+1)] = 0$.
In the second term of $\dT$ for the same reason only $\jo =0$ can contribute,
but then the condition $\jo >\jt$ cannot be satisfied. Similarly neither
term of the volume preserving diffeomorphisms {\scX} survives when operating
on the state $\vert N\rangle$ above, which therefore satisfies the physical
state conditions,
$$
\dT \vert N\rangle = X_{J\cm}^{(+)}\vert N\rangle = 0\ . \equn
$$
It is perhaps noteworthy to remark that even though we have not imposed
the full time reparametrization constraints $\langle N'\vert T_{_{JM}}\vert
N\rangle = 0$ and indeed have argued in the previous sections that these
constraints should  not be imposed {\it a priori}, nevertheless we find
that these
conditions  are satisfied as well for all $J$ {\it except}
$J=0$ which is the Hamiltonian constraint and must be treated separately.

Finally we should apply the Hamiltonian constraint with the correct shift
determined in the last section, namely $H \vert phys\rangle = 4 \vert
phys\rangle$.
Clearly, of all the surviving states $\vert N\rangle$ in {\scN} this selects
only the level $N=4$ state as physical. The significance of this
one surviving physical state is that it is created by the spacetime integral
of the conformal weight $4$ operator, $:\Phi^4:$ acting on the vacuum, $\vert
0\rangle$.  This integrated operator is the {\it only} non-trivial
conformal scalar
one can form in this free field example. Indeed under the conformal
transformation
from flat space to
$R \times S^3$ given by {\cmap}, the volume element
$$
d^4x = e^{-4\t}d\t\, d\O \rightarrow ie^{-4it}dtd\O \equn
$$
after analytic continuation to Lorentzian signature. Hence the integrated
flat space operator $:\Phi^4:$ picks up a factor of $e^{-4it}$ when transformed
to $R \times S^3$ and we find
$$
[H, \int dt d\O e^{-4it} :\Phi^4:] \vert 0\rangle =
-i\int dt d\O e^{-4it}\pa_t :\Phi^4:\vert 0\rangle =
4 \int dt d\O e^{-4it}:\Phi^4:\vert 0\rangle \equn\put\phif
$$
after integration by parts. Since the time integration
in {\phif} selects only the time dependence $e^{4it}$ from the normal ordered
quantity $:\Phi^4:\vert 0\rangle$, we see immediately from the mode
expansion {\fock} that this yields precisely the physical state
$$
\vert phys\rangle = (\varphi^{\dag}_{00})^4 \vert 0\rangle\ .\equn
$$
Hence the integrated operator $\int d^4x :\Phi^4:$ is an explicit
realization in this free scalar theory of a weight $4$ operator $\co_4$
creating a
physical state in {\creat}, as we discussed at the beginning of Section $3$.

The fact that there is only one physical state ({\it i.e.} the ``vacuum")
and only one non-trivial invariant operator surviving the diffeomorphism
constraints
in the scalar $\Phi$ theory is a reflection of the fact that there is no
gravitational dynamics in this theory without a gravitational action.
However, the
correspondence of physical states with operators of dimension $4$ is a general
feature of any diffeomorphic invariant theory. The correspondence of scalar
operators with physical states confirms the interpretation of the shift in the
global Wheeler-DeWitt equation found in Sec. 3, and will be exploited
further in the
study of the effective theory of the conformal factor in paper II.

\vskip 1.5truecm
\centerline{{\it Acknowledgments}}
\vskip .5cm

I.A. and P.O.M. would like to acknowledge the hospitality of
Theoretical Division (T-8) of Los Alamos National Laboratory.
E.M. and P.O.M. would like to thank the Centre de Physique Th\'eorique
of the Ecole Polytechnique for its hospitality. All three authors
wish to acknowledge NATO grant CRG 900636 for partial financial support.

\vfill
\eject

\baselineskip=15pt
{\bf REFERENCES }
\vskip.5cm

\item{[\cFIG]} A. M. Polyakov, {\it Phys. Lett.} {\bf B103} (1981) 207, 211;
\hfill\break\indent{\it Gauge Fields and Strings},  (Harwood Academic, Chur,
1987);\hfill\break\noindent S. K. Blau, Z. Bern, and E. Mottola, {\it Phys.
Rev.}
{\bf D43} (1991) 1212; \hfill\break\noindent P. O. Mazur, {\it Phys. Lett.}
{\bf
B262} (1991) 405;\hfill\break\noindent E. Mottola, {\it Jour. Math. Phys.}
{\bf 36}
No. 5 (1995) 2470.
\hfill\break

\item{[\cAntMot]} R. J. Riegert, {\it Phys. Lett.} {\bf B134} (1984) 56;
\hfill\break\noindent
I. Antoniadis and E. Mottola, {\it Phys. Rev.} {\bf D45} (1992) 2013;
\hfill\break\noindent
S. D. Odintsov, {\it Z. Phys.} {\bf C54} (1992) 531.\hfill\break

\item{[\cAMM]} I. Antoniadis, P. O. Mazur, and E. Mottola, {\it Nucl.
Phys.} {\bf B
388} (1992) 627.\hfill\break

\item{[\cGSW]} M. B. Green, J. H. Schwarz, and E. Witten, {\it Superstring
Theory,
Vol. 1}, \hfill\break\indent (Cambridge Univ. Press, Cambridge,
1987).\hfill\break

\item{[\cCT]} T. L. Curtright and C. B. Thorn, {\it Phys. Rev. Lett.} {\bf 48}
(1982) 1309; \hfill\break\indent {\it Erratum, ibid.}
1768;\hfill\break\noindent
J. L. Gervais and A. Neveu, {\it Nucl. Phys.} {\bf B199} (1982) 59;
\hfill\break\indent
{\bf B209} (1982) 125; {\bf B224} (1983) 329.\hfill\break

\item{[\cBD]} D. G. Boulware and S. Deser, {\it Jour. Math. Phys.} {\bf 8}
(1967) 1468.\hfill\break

\item{[\cWDW]} B. S. DeWitt, {\it Phys. Rev.} {\bf 160} (1967) 1113;
\hfill\break\noindent
J. A. Wheeler, in {\it Battelle Recontres}, C. DeWitt and J. A. Wheeler, eds.
\hfill\break\indent(Benjamin, New York, 1968).\hfill\break

\item{[\cVar]} D. A. Varshalovich, A. N. Moskalev, and V. K. Khersonskii,
{\it Quantum Theory}\hfill\break\indent
{\it of Angular Momentum}, (World Scientific, Singapore, 1988);
\hfill\break\noindent
M. A. Rubin and C. R. Ord\'o\~nez, {\it Jour. Math. Phys.} {\bf 25} (1984)
2888.
\hfill\break

\vfill
\eject

\baselineskip=20pt
\newcount\eqncount
\setbox\sect =\hbox{A}
\beginsection{Appendix}

Because the symmetry group of $S^3$ is $O(4)\cong SU(2) \times SU(2)/Z_2$
the finite dimensional irreducible representations of this group are
labeled by a pair of $SU(2)$ indices $(\jo ,\jt)$ each taking on integer
or half-integer values. The scalar harmonics on $S^3$, denoted here by
$\Y$ are eigenvectors of the scalar Laplacian on the sphere,
$$\eqalign{
-\tri \Y &= (2J)(2J + 2) \Y = 4J (J + 1)\Y \ , \cr
M \equiv (m,m') &= (-J,-J), (-J+1,-J), \ \dots\ ,(J,J-1), (J,J) \cr}
\equn
$$
and belong to the $(J, J)$ representation of $SU(2) \times SU(2)$
with multiplicity $(2J+1)^2$. They form a complete set in which
to expand scalar functions regular on $S^3$,
are mutually orthogonal with respect to integration over $d\O$ and may
be normalized in the standard way:
$$
\int_{S^3} d\O \, Y^*_{_{J'M'}}\Y = \d_{_{J'J}}\d_{_{M'M}}\ .
\equn\puu\Ynorm
$$

Because $S^3$ is the group manifold of $SU(2)$, the
spherical harmonics may also be viewed as rotation matrices that
map the $J$ representation of $SU(2)$ into itself. Hence the $\Y$ are
proportional to the $SU(2)$ Wigner $D$ functions. In fact,
$$
\Y (\a,\b,\g)= \sqrt{2J + 1\over V}\,D^J_{mm'}(\a ,\b ,\g)
\equn\puu\wig
$$
in the Euler angle $(\a, \b,\g)$ parameterization of the three sphere,
$$
d\O^2 = \hbox{$1 \over 4$}(d\a^2 +  d\b^2 + d\g^2 + 2\cos\b\, d\a\, d\g)\ ,
\equn\puu\metric
$$
in which $\a$ and $\g$ have the range $[0, 2\pi]$ and $[0, 4\pi]$
(or {\it vice versa}), and $\b$ has the range $[0, \pi]$. In this
parameterization the volume element on $S^3$ is
$$
d\O = \hbox{$1 \over 8$}d\a\,d\b\,d\g\, \sin\b
\equn
$$
and the volume $V=2\p^2$.

{}From the known properties of the Wigner $D$ functions [\cVar] we obtain
the conjugation relation,
$$\eqalign{
\Ys &= \e_{_{M}} Y_{_{J\ -M}} \qquad {\rm where} \cr
\e_{_{M}} &\equiv (-)^{m-m'}\cr}
\equn
$$
and the integral of three scalar harmonics,
$$
\int_{S^3} d\O\  \Ys Y_{_{\jo\mo}} Y_{_{\jt\mt}} = {\sqrt{(2\jo +
1)(2\jt + 1)\over (2J + 1)\ V}}\, C^{Jm}_{\jo m_1\jt
m_2}C^{Jm'}_{\jo m'_1\jt m'_2}\ ,
\equn
$$
in terms of two ordinary $SU(2)$ Clebsch-Gordon coefficients
$C^{Jm}_{\jo m_1\jt m_2}$, from which {\cgpro} of the text follows
immediately.

The four scalar harmonics at $J=\ha$ lie in the $(\ha,\ha)$
representation of $SU(2) \times SU(2)$ which is the vector
representation of $O(4)$, and play a distinguished role in
the analysis of the diffeomorphism constraints. Because these
four harmonics span the space of translations of the $S^3$
in its embedding $R^4$, the gradients $\o_i^{(M)} =\nabla_i Y_{_{\ha M}}$
satisfy the relations,
$$\eqalign{
\nabla_i\nabla_j Y_{_{\ha M}} &= - g_{ij} Y_{_{\ha M}}\cr
\tri Y_{_{\ha M}} &= \nabla_i \o^{i\,(M)} = - 3 Y_{_{\ha M}}\ ,\cr}
\equn\puu\ckvha
$$
which may be verified directly from the explicit representations
{\wig} and {\metric} as well. Hence the four $\o_i^{(M)}$
are the four proper conformal Killing vectors of $S^3$,
$$
\nabla_i \o_j^{(M)} + \nabla_j \o_i^{(M)} = \hbox{$2 \over 3$} g_{ij}
\nabla_k\o^{k\,(M)}\ ,
\equn\puu\ckvapp
$$
and generate four special volume non-preserving coordinate transformations
of the sphere. From this relation and the definition {\pcv}
one immediately derives {\divpcv} of the text and the fact that
$\k_{_M}^{(\pm)}$ satisfy the conformal Killing equation of the full
metric on $R \times S^3$. Furthermore, using {\ckvha} we observe that
$$
\nabla_j\Bigl( \nabla_i Y_{_{\ha M'}}^* \nabla^i Y_{_{\ha M}}\Bigr) =
\nabla_j \Bigl(- Y_{_{\ha M'}}^*Y_{_{\ha M}}\Bigr)\equn
$$
so that the quantities in parentheses are equal up to a constant.
The constant may be evaluated by taking the volume integral of each
on $S^3$. In this way, we find
$$
(\nabla_i Y_{_{\ha M'}}^*) (\nabla^i Y_{_{\ha M}}) + Y_{_{\ha M'}}^*Y_{_{\ha
M}}
= \hbox{$4 \over V$} \d_{_{\mo\mt}}\d_{_{\mo'\mt'}}\ .\equn\puu\ysum
$$
The six Killing vectors $\rho^i$ of the sphere may also be expressed in terms
of these same $J=\ha$ scalar harmonics by eq. {\ks} of the text. Using
this definition, the definition of the Lie bracket contained in
eqs. {\spat} and {\lbrdef} and the relation {\ysum}, it is straightforward
to derive the classical Lie algebra {\liebr} of the conformal group
of the Einstein universe $R \times S^3$.

The six Killing vectors $\rho^i$ are actually the first non-trivial
representatives
of the transverse vector harmonic functions on the 3-sphere, which we denote
generally by $\cy^i_{J\cm}$. These belong to the
$(J + \ha, J-\ha)$ or $(J-\ha, J+ \ha)$ representations
of $SU(2) \times SU(2)$, so that the magnetic index $\cm$ ranges over
$2\times(2J+2)(2J)= 8J(J+1)$ values and $J\ge \ha$. The vector
harmonics are transverse,
$$
\nabla_i\cy^i_{_{J\cm}} = 0 \equn
$$
and are eigenfunctions of the Laplacian on $S^3$,
$$
-\tri \cy^i_{_{J\cm}} = \left[4J(J+1) - 1\right]\cy^i_{_{J\cm}}\ .
\equn\puu\lapvec
$$
At $J=\ha$ the $(1,0)$ and $(0,1)$ representations give
$3+3=6$ transverse vector harmonics which are precisely the
$6$ Killing vectors of $S^3$,
$$
\nabla_i \rho_j + \nabla_j\rho_i = 0\ ,
\equn
$$
and take the special form {\ks}. In this case the index $\cm$ takes on
the six values corresponding to the adjoint
representation of $O(4)$, labeled by an anti-symmetrized pair of
four-vector indices
$(M_1,M_2)$. None of the higher $J > \ha$ vector harmonics can be written as
simply in terms of products of scalar harmonics and their gradients. Taking the
divergence of the Killing equation and using
$[\nabla_j,\nabla_i]\rho^j = R_{ji}\rho^j = 2 \rho^i$ on the unit sphere
gives
$$
\tri \rho_i + 2\rho_i = 0\equn
$$
which agrees with the eigenvalue equation {\lapvec} for the $J=\ha$ vector
harmonics [\cVar].

All of the vector harmonics $\cy^i_{_{J\cm}}$ are mutually
orthogonal and may be normalized in the standard way, {\it i.e.}
$$
\int_{S^3} d\O \, \cy^{i*}_{_{J'\cm '}}\cy_{i\, _{J\cm}}= \d_{_{J'J}}\d_{_{\cm
'\cm}}\ .
\equn\puu\cynorm
$$
Together they form a complete basis for the volume preserving diffeomorphisms
of the sphere. The Killing vectors defined by {\ks} are normalized
differently, in order to satisfy the classical Lie algebra {\liebr} and obey
$$
\int_{S^3} d\O \, \rho^{i*}_{_{M_1M_2}}\rho_{i_{M_1M_2}} = \hbox{$V \over
2$}\left(\d_{_{M_1M_3}}\d_{_{M_2M_4}} -
\e_{_{M_1}}\e_{_{M_2}}\d_{_{M_1\,-M_4}}\d_{_{M_2\,-M_3}}\right)
\equn\puu\kvnorm
$$
instead of {\cynorm}. Finally, in deriving the quantum realization of the
classical conformal algebra {\liebr} one also encounters
$$\eqalign{
\int_{S^3} d\O \, \rho^i_{_{M_1M_2}}&\rho^{j*}_{_{M_3M_4}}
\nabla_i\,\rho_{j_{M_5M_6}} =\cr
\hbox{$iV \over 4$} &\Bigl\{\d_{_{M_4M_6}}
\left(\d_{_{M_1M_3}}\d_{_{M_2M_5}} - \e_{_{M_1}}\e_{_{M_2}}
\d_{_{-M_1M_5}}\d_{_{-M_2M_3}}\right)\cr
 &+ \e_{_{M_1}}\e_{_{M_3}}\d_{_{-M_3M_6}}\left(\d_{_{-M_1M_5}}\d_{_{M_2M_4}}
- \d_{_{-M_1M_4}}\d_{_{M_2M_5}}\right)\cr
& - \d_{_{M_4M_5}}\e_{_{M_5}}\e_{_{M_6}}
\left(\d_{_{M_1M_3}}\d_{_{M_2M_6}} - \e_{_{M_1}}\e_{_{M_2}}
\d_{_{-M_1M_6}}\d_{_{-M_2M_3}}\right)\cr
&- \e_{_{M_1}}\e_{_{M_3}}\e_{_{M_5}}\e_{_{M_6}}\d_{_{-M_3M_5}}
\left(\d_{_{-M_1M_6}}\d_{_{M_2M_4}} - \d_{_{-M_1M_4}}\d_{_{M_2M_6}}\right)
\Bigr\}\ ,\cr}
\equn
$$
which is evaluated by repeated use of the relations {\ckvha} and {\ysum}
above.
\end